\documentclass[11pt]{article}
\usepackage{moriond2000,epsfig}

\def\gs{\mathrel{\raise0.35ex\hbox{$\scriptstyle >$}\kern-0.6em
\lower0.40ex\hbox{{$\scriptstyle \sim$}}}}

\def\be{\begin{equation}}
\def\ee{\end{equation}}
\def\bea{\begin{eqnarray}}
\def\eea{\end{eqnarray}}

\begin{document}
\vspace*{4cm}
\title{SHEDDING LIGHT ON THE DARK SIDE OF GALAXY FORMATION:\\
       SUBMILLIMETRE SURVEYS THROUGH LENSING CLUSTERS}

\author{ROB IVISON\,\,\footnote{This work is being undertaken with
Andrew Blain, James Dunlop, Thomas Gr\`eve, Jean-Paul Kneib, Kirsten
Knudsen, John Peacock, Ian Smail and Paul van der Werf.}}

\address{Dept of Physics \& Astronomy, University College London,
Gower Street, London WC1E 6BT, UK}

\maketitle\abstracts{
Most new sub-mm/mm surveys, both deep and shallow, are being targeted
at rich cluster fields. I explain why, comparing surveys that have
exploited weak lensing by massive foreground clusters with those done
in blank fields.}

\section{Historical perspective}

Sub-mm/mm surveys have revolutionised our understanding of star
formation in the early Universe\,\cite{bl99} through the discovery of
a vast population of very luminous
galaxies,\cite{sm97,ba98,hu98,ea99,be00} clarifying the relative
importance of obscured and unobscured emission. Many are extremely
red\,\cite{de99,sm99,idsd} (a factor $\ge$100 in flux between 1 and
2\,$\mu$m) and most are optically invisible, $BV\!RI > 26$, even to
the {\em Hubble Space Telescope}\,\cite{hu98,sm98} ({\em HST}).

The impact of sub-mm/mm surveys has been due to the commissioning of
revolutionary bolometer cameras such as SCUBA\,\cite{ho99} on the
James Clerk Maxwell Telescope and MAMBO\,\cite{kr99} at the Institut
de Radioastronomie Millim\'etrique and the sensitivity of those
devices to heavily extinguished galaxies\,\cite{bl93} -- to {\em `the
optically dark side of galaxy formation'}.\,\cite{gu97} SCUBA, in
particular, has made a huge impact in cosmology through its ability to
measure the bolometric output of $1<z<5$ dust-enshrouded galaxies
(albeit with a resolution of only 14$''$) whose energy distributions
peak in the sub-mm band.

The foundations of sub-mm/mm cosmology are already in place, only a
few years after the commissioning of SCUBA, and the community is
moving rapidly to build on them, developing new telescopes and
instrumentation (e.g.\ the Atacama Large Millimeter Array in Chile,
the Large Millimeter Telescope in Mexico, and the next-generation of
ground-based bolometer cameras, SCUBA-2 and BOLOCAM).

\section{Sub-mm/mm surveys and the nature of sub-mm galaxies}

The first generation of sub-mm/mm surveys, completed and ongoing, are
listed in Table~1.

\begin{table}
\caption{Published and ongoing sub-mm/mm surveys and their claimed
areas and rms sensitivities.}
\vspace{0.4cm}
\begin{center}
\begin{tabular}{|l|c|c|c|}
\hline
                &                    &             &                   \\
Survey name     & Wavelength$^{\rm d}$/ & Area        & Depth (rms)       \\
                & {\sc fwhm} of beam & /arcmin$^2$ & /mJy\,beam$^{-1}$ \\
                &                    &             &                   \\
\hline
                &                    &             &                   \\
{\em Completed:}      &                    &             &                   \\
SCUBA lens survey\,\cite{sm97}&850$\mu$m/14$''$$^{\rm b}$&36$^{\rm b}$&1.7$^{\rm b}$\\
Hawaii survey fields\,\cite{ba99}&850$\mu$m/14$''$&104$^{\rm e}$&2.7$^{\rm e}$\\
HDF\,\cite{hu98} (UK sub-mm survey consortium)&850$\mu$m/14$''$  &5.6  &0.5\\
Hawaii HFF radio-selected survey\,\cite{ba00}   &850$\mu$m/14$''$ &31&2\\
Canada-UK deep sub-mm survey\,\cite{ea99} (CUDSS)&850$\mu$m/20$''$$^{\rm a}$&92&1.2\\
Dutch lens survey\,\cite{pvdw}&850$\mu$m/14$''$$^{\rm b}$&50$^{\rm b}$&$\sim$2$^{\rm b,f}$\\
Canada HFF survey\,\cite{bo01} &850$\mu$m/17$''$  &121&$\sim$3\\
Canada lens survey\,\cite{ch01}  &850$\mu$m/14$''$$^{\rm b}$&42$^{\rm b}$&$\sim$2$^{\rm b}$\\
MAMBO survey\,\cite{be01}&1250$\mu$m/10$''$ &450$^{\rm c}$&0.5$^{\rm c}$\\
{\em Ongoing:}&                &             &                   \\
8mJy survey (UK sub-mm survey consortium)&850$\mu$m/14$''$  &240&2.7\\
High-$z$ signpost survey\,\cite{idsd}&850$\mu$m/14$''$&78&1\\
UK shallow lens survey&850$\mu$m/14$''$$^{\rm b}$&45$^{\rm b}$&2.5$^{\rm b}$\\
A370/A2218 SCUBA lens surveys&850$\mu$m/14$''$$^{\rm b}$&11$^{\rm b}$&0.5$^{\rm b}$\\
A2218 MAMBO lens survey&1250$\mu$m/10$''$$^{\rm b}$&20$^{\rm b}$&0.3$^{\rm b}$\\
                &                    &             &                   \\
\hline
\end{tabular}
\end{center}

{\scriptsize

$^{\rm a}$ Effective {\sc fwhm} is 20$''$ after convolving with beam to
achieve depth of 1.2\,mJy\,beam$^{-1}$ rms.

$^{\rm b}$ Divide values by $\sim$2.5 to calculate the effective
source-plane area/depth/resolution.

$^{\rm c}$ Equivalent to $\sim$1.2\,mJy\,beam$^{-1}$ rms at
850\,$\mu$m for $z \sim 2.5$.

$^{\rm d}$ Note that 450-$\mu$m source counts have also been
reported\,\cite{bl00}.

$^{\rm e}$ Sub-area of 7.7\,arcmin$^2$ to 0.8\,mJy\,beam$^{-1}$ rms.

$^{\rm f}$ Two/two/four fields to 1.5/2/3\,mJy\,beam$^{-1}$ rms.

}
\end{table}

It is apparent that conventional blank fields have soaked up most of
the time spent on cosmology surveys. Areas and rms depths range from
the UKSSC 8-mJy survey's 200\,arcmin$^2$/2.7\,mJy beam$^{-1}$ to the
UKSSC HDF\,\cite{hu98} survey's 5.6\,arcmin$^2$/0.5\,mJy\,beam$^{-1}$,
and MAMBO has now completed its first deep 1250-$\mu$m
survey\,\cite{be01} (450\,arcmin$^2$/0.5\,mJy\,beam$^{-1}$, {\sc fwhm}
10$''$).

These blank-field surveys have been tremendously successful,
determining the 850-$\mu$m source counts above 2\,mJy and thereby
resolving directly up to about half of the {\em COBE} background at 
850\,$\mu$m. The deepest map, of
the HDF\,\cite{hu98}, has also yielded a statistical detection of
the sub-mm emission from Lyman-break galaxies\,\cite{pe00}.

After initial uncertainty, there is now a growing consensus amongst
the sub-mm/mm community that the sources uncovered by SCUBA (and now
MAMBO) are massive, intensely star-forming galaxies at \={\em z} $\sim
3$ (possibly slightly closer\,\cite{li99}), resembling ultraluminous
{\em IRAS} galaxies in some respects, usually with only a tiny
fraction ($<$1\%) of their luminosity released in the rest-frame
UV\,\cite{sa00} (c.f.\,\cite{ad00,la00}) so that many qualify as
`extremely red objects'\,\cite{de99,sm99,idsd} (EROs, $R-K \gs 6$).

The road to this consensus has been paved by painstaking efforts to
determine the nature of individual galaxies, largely through a process
of trial and error, slowly determining the most efficient techniques
for identifying near-IR or optical counterparts, investigating basic
properties and, in pitifully few cases, measuring
redshifts\,\cite{iv98,iv00}.

To date, deep imaging in the radio and near-IR bands\,\cite{sm99,sm00}
have been far and away the most effective techniques, pinpointing
counterparts (see Figures~1 and 2 and their captions) and facilitating
spectroscopic follow up. This has culminated in several CO detections
that suggest molecular gas masses consistent with the formation of
elliptical galaxies.\cite{fr98,fr99,kn00}

Radio flux measurements or limits at 1.4\,GHz have also provided a
plausible redshift distribution\,\cite{sm00} based on new photometric
techniques\,\cite{cy99,cy00}. Other techniques -- mm interferometry,
for example\,\cite{do99,be00,ge01} -- have been less successful at
elucidating what we know of the SCUBA galaxy population, but clearly
hold promise for the future\,\cite{fr00}, particularly for very bright
sources ($\gs$8\,mJy at 850\,$\mu$m) found in the field, through
cluster lenses or near luminous radio galaxies\,\cite{idsd}. There are
hopes that broad-band spectral devices may be able to determine
spectroscopic redshifts using CO transitions, regardless of the
availability of plausible optical/IR counterparts, though the
technical challenges are immense.

\begin{figure}[p]
\psfig{figure=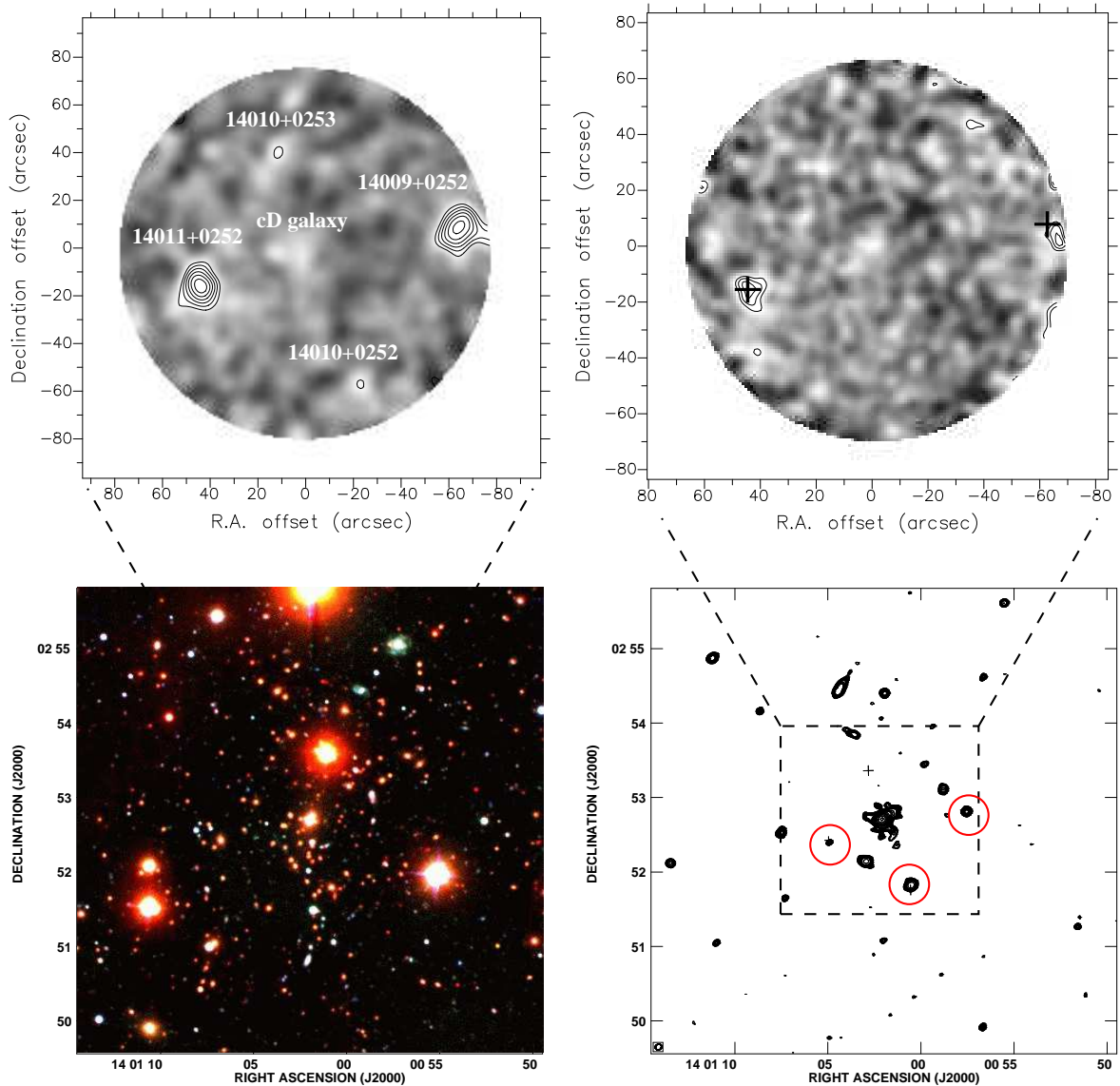, width=160mm}
\caption{An illustration of the detection and further investigation of
sub-mm galaxies in the Abell\,1835 cluster field$^{30}$. {\em Top
left}: 850-$\mu$m map (field diameter $\sim150''$); {\em top right}:
450-$\mu$m map. Sources are labelled on the 850-$\mu$m image. Below
these are a true-colour $UBI$ image ({\em lower left}) and a 1.4-GHz
map from the VLA ({\em lower right}), both $\sim370''$ across. The
sub-mm sources are easily detected in the very deep radio map
(red circles), with far better positional accuracy than afforded by
SCUBA. Twenty other radio sources seen in the VLA image can be used to
co-align the radio/optical (or radio/near-IR) coordinate frames,
yielding counterpart positions accurate to 0.1$''$. Note that the
radio image shown here represents only 6\% of the VLA's primary beam
area at 1.4\,GHz.}
\end{figure}

\begin{figure}
\psfig{figure=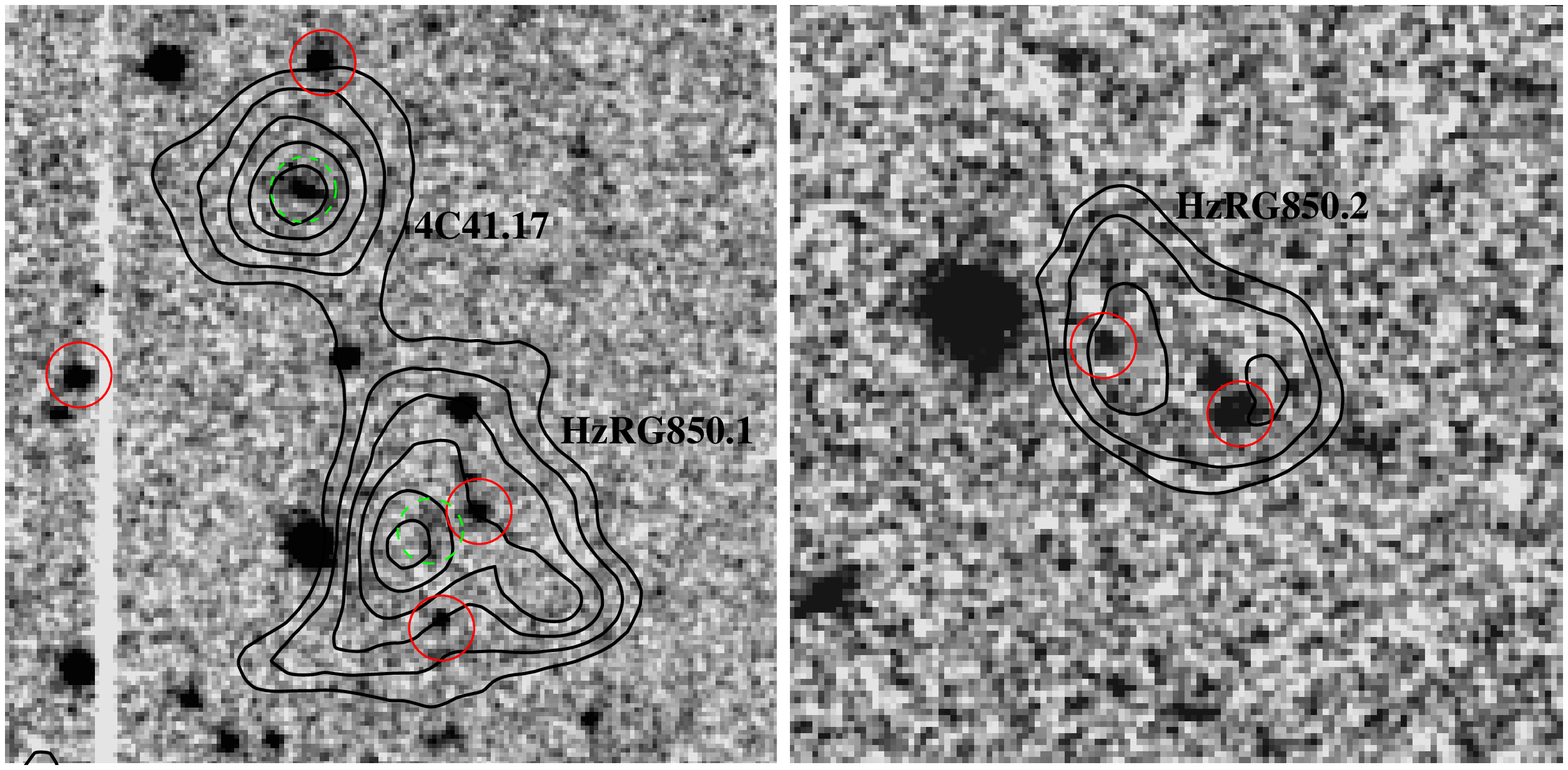, width=160mm}
\caption{$K$-band images of regions around the $z=3.8$ radio galaxy,
4C\,41.17, where there is an order-of-magnitude over-density of sub-mm
sources$^{31}$. 850-$\mu$m data are shown as contours. Red circles
(solid lines) denote EROs ($R-K \gs 6$). Two EROs are probably
associated with the blended sub-mm galaxy, HzRG850.2. For the other
sub-mm galaxies (HzRG850.1, 4C\,41.17), green circles (dashed lines)
denote the likeliest counterparts (faint and red, in the case of
HzRG850.1, though a bona fide ERO could also be responsible for the
sub-mm emission).}
\end{figure}

The current samples of sub-mm/mm galaxies contain a small but
significant fraction of active galactic nuclei (AGN), though deep,
hard-X-ray imaging\,\cite{fa00,ho00,mu00} has so far failed to uncover
the large, heavily obscured AGN population that some had suspected
from the earliest follow-up work\,\cite{iv98} and from theoretical
arguments\,\cite{al99}.

\section{The problem of confusion -- lifting and separating with a lens}

Had the galaxies discovered in sub-mm surveys been only fractionally
fainter or less numerous, a second, more sensitive generation of
bolometer cameras would have been required to discover them. Early
surveys\,\cite{sm97,ba98,hu98,ea99} would collectively have uncovered
only a couple of sources -- the first, an obvious AGN\,\cite{iv98}
(SMM\,J02399$-$0136) and the second,\cite{hu98} a puzzle with no
optical or near-IR counterpart (HDF850.1). Who can say what
conclusions might have been reached and how future surveys, e.g.\ with
{\em FIRST}, may have suffered?

We have been fortunate, then, that first-generation bolometer arrays
were sufficiently sensitive to enable rapid progress in sub-mm
cosmology. We have been less fortunate regarding source confusion: few
would have predicted that SCUBA would reach its effective confusion
limit\,\cite{bl98} within a few months of being commissioned. The
deepest direct counts\,\cite{hu98} are already at the confusion limit,
suggesting that further progress in constraining the intensity of the
sub-mm background and the nature of the faint sub-mm population
requires an innovative approach.

To probe below the confusion limit using the existing sub-mm/mm
cameras requires the use of the natural magnifying glasses that
provide the raison d'$\hat{\rm e}$tre for this conference:
gravitational lenses. Massive clusters provide a magnified (although
distorted) image of a small region of the background sky; thus both
the effective resolution and sensitivity of the survey are increased,
as measured on the background sky. This enables surveys to probe faint
flux densities without suffering confusion, albeit at the price of a
distorted view.

With an accurate cluster mass model, the distortion can be
corrected. The first lens survey\,\cite{sm97,bl99} illustrated the
advantages of this approach for the counts.\cite{deep} About 100\% of
the {\em COBE} 850-$\mu$m background was resolved down to
0.5\,mJy. Follow-up
observations,\cite{fr98,iv98,fr99,sm99,fr00,iv00,sm00} also benefitted
from achromatic gravitational amplification: not only was the
effective depth of the sub-mm maps increased, but the counterparts at
all other wavelengths were similarly amplified. This allows useful
follow-up observations to be obtained in several hours or tens of
hours using the current generation of telescopes and instrumentation:
it is no coincidence that of the $\sim$100 known sub-mm galaxies, only
a handful have reliable spectroscopic redshifts and {\em all} of these
were discovered through cluster lenses.

Another advantage of using clusters is that extraordinarily deep
images -- X-ray, optical, IR and radio -- exist or are scheduled for
these fields. The HDF is the only blank field that is equally
blessed. Abell 851, 1835 and 2218 (and many other cluster fields) have
superb {\em HST} images and near-IR data; Abell 370, 851 and 2125 have
1.4-GHz maps with $<$10-$\mu$Jy\,beam$^{-1}$ noise levels.

\section{Future plans and concluding remarks}

Following on from the success of the earliest sub-mm cluster
survey\,\cite{sm97,bl99}, groups in the UK, Holland and Hawaii are
currently undertaking more surveys with SCUBA and MAMBO that exploit
cluster lenses. The latest of these will combine a long integration
(equal to that obtained on the HDF) with amplification by the cores of
amongst the most massive, well-constrained cluster lenses known,
A\,370 and A\,2218.

At modest amplifications ($A\sim$\,2--5), it should be possible to
detect the optically-identified arclet population; probing fainter, it
is likely that a new, largely unexplored class of lensed feature may
appear: multiply-imaged pairs, recognised in the first {\em HST}
cluster images.

These appear in the optical/near-IR as symmetric images with typical
separations of 5--10$''$ (i.e.\ within a single SCUBA or MAMBO beam)
and can be simply and successfully modelled as highly magnified images
($A \sim$\,10--100) of very faint, compact sources which lie close to
a critical line. In a well-constrained lens such as A\,2218, their
location in the cluster, combined with the positions and separation of
any radio/IR/optical counterparts, can give the source redshift and
amplification to high precision. The area of the source plane in which
pairs are formed can also be estimated from the lens models, allowing
their rate of occurence to be converted into an estimate of the
surface density of extremely faint (tens of $\mu$Jy) sub-mm/mm
background sources, with the bonus of crude redshift information.

Using the superb recent {\em HST} imaging of A\,2218, at least 4
highly magnified pairs have been identified (from a source population
with a comparable surface density to that expected for the very faint
sub-mm/mm population, $\sim$10\,arcmin$^{-2}$) suggesting that the
cluster amplification cross-section is high and that the chance of
finding such systems is good. Failure to detect any of these highly
magnified sources using SCUBA and MAMBO would indicate convergence of
the source counts and can be used to impose strong limits on the
surface density of very faint sources and the total intensity in
resolved sources in the sub-mm background.

\section*{Acknowledgments}
I acknowledge a PPARC Advanced Fellowship and support from the
Training and Mobility of Researchers Programme.

\end{document}